\documentclass[twocolumn,%
reprint,
 amsmath,amssymb,aps,
floatfix
]{revtex4-2}

\usepackage{graphicx}
\usepackage{dcolumn}
\usepackage{bm}
\usepackage{float}

\usepackage{xcolor}

\begin{document}


\title{ Crystallization in single and multicomponent Neutron Star  crusts.}

\author{D. Barba González$^1$}
 \email{david.barbag@usal.es}
\author{ C. Albertus}
 \email{albertus@usal.es}
\author{M. A. Pérez-García.}
 \email{mperezga@usal.es}

\address{Department of Fundamental Physics and IUFFyM, Universidad de Salamanca,Plaza de la Merced s/n E-37008, 
Salamanca, Spain\\
}

\begin{abstract}
We use Molecular Dynamics simulations to study the formation and stability of single and multicomponent lattices in the outer crust of Neutron Stars. Including an improved treatment for Gaussian charge distributions of ions we obtain the expressions for the potential and forces arising from electron screened Coulomb interactions using the efficient Ewald sum procedure. 
Our findings show that for baryon densities in the outer crust,  a point-like ion treatment can not fully describe the crystallization behaviour. In our work, the usual Coulomb parameter, $\Gamma_C$, along with the screening parameter from an electron polarizable background, $\kappa$, are complemented with an additional parameter, $\eta$, providing information on the finite size of ions. In our approach we find that including beyond point-like approaches in screened ion plasmas under the Thomas-Fermi approximation has a  strong impact on calculated lattice energetic stability  decreasing crystallization energies per baryon up to $\sim 40\%$ with respect to point-like interaction and, as a consequence, melting point resulting displaced to lower temperatures.
\end{abstract}

\keywords{screening, one component plasma, outer crust, neutron star}

\maketitle

\section{Introduction}

Neutron Stars (NSs) are compact stars that appear at the end-point of the evolutionary path of progenitor stars with masses larger than about $\sim8 M_\odot$. These collapsed stars display a rich internal structure with matter densities spanning many orders of magnitude from the core to the outer crust. Some of the complexity related to the microscopic modelling of their interior relies on the fact that matter must be described under different Lorentz regimes, from the relativistic nature of fluid interacting matter in the core to the nearly classical ion dynamics in the outer crust. Quantum effects arising from relativistic leptons, superconducting and superfluid components \cite{Chamel_2008} and correlated states \cite{Ivanytskyi_2019} are also present.

In the crust, thermodynamical quantities are intimately dependent on the composition and the microscopic model of matter interaction. In this line, there is recent interest in studying multicomponent systems arising from fully or partially accreted crusts \cite{suleiman22}. This originates from NS binary systems hosting the process where matter from the companion star is intermittently transferred to the NS surface and processed producing highly luminous bursts in X-rays. Observing the cooling of soft quiescent X-ray transients allows to probe our understanding of NS composition from remaining ashes \cite{Mckinven_2016}, arrangement \cite{Medin_2010,Horo_2007,CaplanMulticomponent} and the deep crustal heating mechanism \cite{Brown_1998}. Let us remind here that the composition can impact not only the generic equation of state (EOS) i.e. the relation pressure versus energy density at a given temperature, $p(\varepsilon, T)$ but specifically crust  properties such as the thermal conductivity and shear modulus with important consequences on the cooling of the star, the evolution of the magnetic field or the seismic activity as described in \cite{PotekhinBook, fantina18,suleiman22,gusakov, Samuelsson_2007} just to cite some previous works.

It is now well established that in the outer NS crust  matter arranges  itself into periodic structures. An effective description of these very outer layers based on low density ionic matter screened by electrons seems thus appropriate. As established, for crustal densities beyond $10^{6}$ $\rm g/cm^3$ electrons are in the degenerate relativistic regime. It is in this scenario that a first approach to the complex picture emerges invoking the presence of Coulomb crystals, in which atoms are fully ionized and electrons form a neutralizing (polarizable) fluid surrounding the periodic ion array. It has been thoroughly studied in the literature, see  \cite{potekhinchabrier_PhysRevE.62.8554, BaikoPhysRevE.66.056405,Haenselbook} and references therein. 

{One particular realization is the one component plasma (OCP), an idealized neutral plasma with a single species of ions of charge $Z$ (in units of electron charge $e$), mass $m_I$ and ion number density $n_I$ (or equivalently $\rho$ mass density) where the cold electron fluid is described by the relativistic parameter $x_{\mathrm{r}} \equiv \frac{p_{F,e}}{m_e} =\frac{\left(3 \pi^2 Z n_I\right)^{1 / 3}}{m_e} \approx 0.01\left(\rho \frac{Z}{m_I}\right)^{1 / 3} \gg 1$. $p_{F,e},m_e$ are the electron Fermi momentum and mass, respectively (we use $\hbar=c=1$).

A series of works have calculated the effect of electron exchange correlations or polarizability, mostly within the linear response formalism, see for example \cite{PollockHansen_PhysRevA.8.3110, potekhinchabrier_PhysRevE.62.8554, KozhberovPRE103}.
These calculations mainly focus, in the static regime, on the electron longitudinal dielectric function $\epsilon(k)$ dependent on momentum $k$ in Fourier space. In the present work we will assume that the full Debye screening at finite ion temperature i.e.  $T\equiv T_I$, is  characterized by  $k_D^{-1}=\lambda_D\approx k_{TF}^{-1}$, approximately that due to cold degenerate electrons. For a system of ions in the  polarizable electron background it can be written \cite{KozhberovPRE103}, to first order corrections, under the form $\epsilon(k)=1+\frac{k^2_{TF}}{k^2} \epsilon_2(k)$ with $k_{\mathrm{TF}}=\left(4 \pi e^2 \partial n_e / \partial \mu_e \right)^{1 / 2}$  the Thomas-Fermi (TF) wave number and  $n_e$, $\mu_{\mathrm{e}}=m_{\mathrm{e}} \sqrt{1+x_{\mathrm{r}}^2}$ being the electron number density and chemical potential, respectively. 

Studies in Earth laboratories provide information on the mass, charge and stability of nuclei  \cite{Wang_2021}. Some of the known neutron rich isotopes are expected to appear in this NS environment. Typical values of ion density in the outer crust are below the neutron drip density $\rho_{ND}\sim 4 \times 10^{11}$  $\rm g/cm^3$ such that the screening parameter $k_{TF}l \lesssim 1$, where $l=n_I^{-1 / 3}/\xi$  is the mean interparticle distance and $\xi=(4 \pi/3)^{1 / 3}$. Under these conditions the linear response approximation can account for electron polarization to describe the system. At slightly higher densities the neutron gas may distort lattice stability \cite{Kobyakov_2014}.

Different approximations for the responsive electron sea are encoded in the form of $\epsilon_2(k)$ as we will discuss in what follows. Using \cite{KozhberovPRE103},  $\epsilon_2(k)=0$ corresponds to the case of a rigid background, being finite in the polarizable case. Typically, the relativistic approach of Jancovici \cite{jancovici} for degenerate electrons recovers, in the non-relativistic limit $x_r \ll1$, the familiar Lindhard form, while for small momentum $k\ll 2 k_{F,e}$ it yields the widely used Thomas-Fermi approximation $\epsilon_2(k)=1$. We must note that although the Thomas-Fermi approach is widely used in ion dynamics, even in the context of previous NS crust calculations \cite{CaplanMulticomponent}, the reliable polarizable electron background is recovered from a more general, relativistic treatment such as that from RPA by Jancovici \cite{jancovici}. The reason for this is  the fact that at larger values of the screening parameter $\kappa\equiv k_{TF}l$  the linear response approach fails. At larger $x_r$ the screening in the ultrarelativistic
electron gas is only determined by the ion
charge number $Z$ being this behaviour not captured by the simplified TF model \cite{BaikoPhysRevE.66.056405}.

Calculations using Monte Carlo and Molecular dynamics (MD)  simulations \cite{Meijer, hamaguchi, CaplanMulticomponent} with a prescribed Yukawa potential are consistent with a TF polarizable electron gas. However, more refined estimates of electrostatic energy in Coulomb systems \cite{KozhberovPRE103} find structural differences in the  lattices when TF or Jancovici models are used, finding bcc lattices as the ground state for systems with $k_{TF}l<1.0657$. Although the TF model has been widely used, calculations using Jancovici models in degenerate electron backgrounds can yield significant  electrostatic energy corrections for small $Z$, see Fig.1 in that same ref. \cite{KozhberovPRE103}.

In this work we are interested in exploring additional energetic corrections derived within the TF approximation but including beyond point-like approaches both in OCP and multicomponent plasmas (MCP). In this scenario, species population at low temperatures has been calculated in \cite{pearson2018, Fantina_2020} being  most likely an ion  distribution peaking at given baryon mass $A$ with some spread, see \cite{CaplHorlambda}. 

This manuscript is organized as follows. In Sec. \ref{sect2} we discuss our formalism,  including the  Molecular Dynamics setup used and the efficient procedure for energetics and force calculation using the Ewald sum for screened OCP and MCP systems with Gaussian finite size charge spread. In Sec. \ref{results} we present our results arising from our computational simulations and discuss our findings in light of other existing calculations. Finally in Sec. \ref{conclude} we summarize and conclude. }

\section{Screened OCP and MCP system with Gaussian charge distribution}
 \label{sect2}
 We aim to describe in detail the screened OCP or MCP, a charge-neutral system with ion density $n_I=\sum_{i} X_i n_{I,i}$, composed of $i$ ion types, each carrying electric charge $Z_i$ and mass number $A_i$ in the sample of $N_I$ ions. Number fractions are defined as $X_i=N_{I,i}/N_I$.  As mentioned, at crustal densities beyond $10^{6}$ $\rm g/cm^3$  electrons have been stripped off the atoms forming a relativistic degenerate Fermi sea, thus the potential created at distance $r$ by the ith ion at position $\vec{r_i}$ is not the nude Coulomb but for the screening conditions considered, where $k_{TF}l \lesssim 1$, it will be approximated in the linear response by the so-called {\it Debye potential}.

\begin{figure}[t]
\centering{\includegraphics[width=0.97\linewidth]{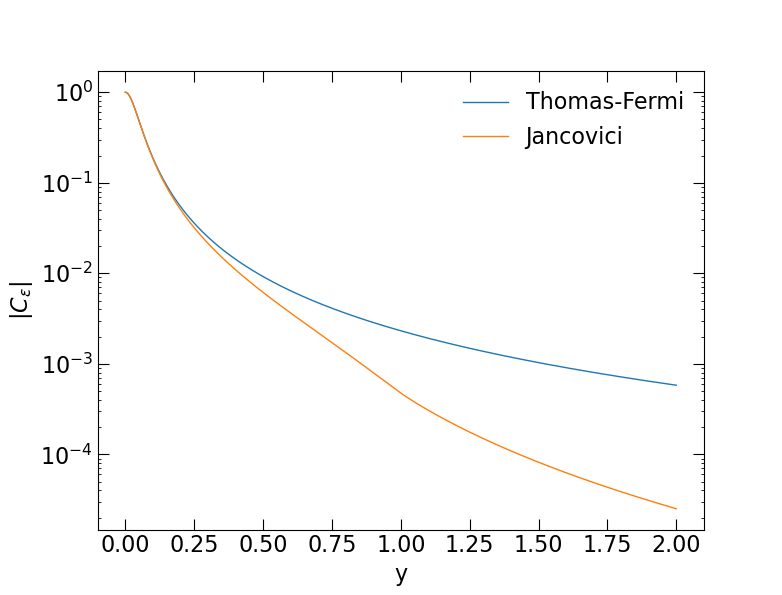}}
\caption{Polarization coefficient $|C_\epsilon|\equiv |\epsilon^{-1}-1|$ as a function of $y=k/2 k_{F,e}$ in the electron polarization energy \cite{BaikoPhysRevE.66.056405} in the Thomas-Fermi and Jancovici \citep{jancovici} approximations. See text for details. }
\label{figcoeff}
\end{figure}

In terms of the static dielectric  function $\epsilon(k)$ the energetic correction due to the electron polarization response \cite{BaikoPhysRevE.66.056405}  will depend on $\epsilon^{-1}(k)-1$ so that for $y\equiv k/2 k_{F,e} \gtrsim 1/4$ a clear departure arise among TF and Jancovici models \cite{Haenselbook}, see Fig.(\ref{figcoeff}), involving important consequences, such as  a robustly less bound bcc lattice in the ground state for the relativistic model. Note that this limit of large $y$ corresponds to typical small distances $1/r\sim k \sim y$.
 
In the TF approximation for the dielectric function in Fourier space, the space dependent potential displays a Yukawa-like form
\begin{equation}
    \phi_i\left(\vec{r}\right)=\frac{Z_i}{|\vec{r}-\vec{r_i}|}e^{-\frac{|\vec{r}-\vec{r_i}|}{\lambda_e}},
    \label{phi1}
\end{equation}

{ where $\lambda_e\equiv \lambda_{TF}=k^{-1}_{TF}$ is the TF screening length. In the relativistic limit it fulfills 
$k_{\mathrm{TF}} l \approx 0.185 Z^{1 / 3} \frac{\left(1+x_{\mathrm{r}}^2\right)^{1 / 4}}{x_{\mathrm{r}}^{1 / 2}}$ for degenerate electrons. Electron number density is parameterized in the charge neutral system as $n_e=\sum_i Z_i n_{I,i}=\frac{k^3_{F,e}}{3 \pi^2}$.

 At this point it is worth discussing how the thermodynamical quantities of interest will be calculated from the crystal/fluid  configurations found in equilibrium as dictated by Molecular Dynamics in an interacting system of ions in presence of a polarizable electron background. In this setup, widely used in many-body systems, see for example \cite{farouki&hamaguchi, honeycutt} for Yukawa or Lenard-Jones systems, we solve the equations of motion for ions considering dynamical oscillations will be much smaller, $\omega \ll \omega_p$, than their associated  plasma frequency, $\omega_{\mathrm{p}}=\left(4 \pi e^2 n_{\mathrm{I}} Z^2 / m_{\mathrm{I}}\right)^{1 / 2}$, so that the screening is instantaneous and the dielectric function $\epsilon(k, \omega)\simeq \epsilon(k)$ only depends on momentum $k$. This fact is responsible for  modifying the effective forces (and dynamics) in the screened ion system.
 
 Thus in our treatment we will not follow the dynamics of relativistic electrons in the charge neutralizing gas, being the screening of the ionic Coulomb interaction the result of the responsive electron gas. Ion positions and momenta, even in crystallized states, are used to obtain subsequent magnitudes resulting from the simulation after reaching long-term stable equilibrium starting from the initially randomized ion phase space at the designated NVT ensemble.
 
 As stated in \cite{PotekhinBook} thermal de Broglie wavelengths of free ions $\lambda_{\mathrm{dB,I}}=\left(\frac{2 \pi }{m_I k_{\mathrm{B}} T}\right)^{1 / 2}$ allow sizing the importance of quantum effects on their motion, i.e. when $\lambda_{\mathrm{I}} \gtrsim l$ or at $T \ll T_{\mathrm{p}}$, where $T_{\mathrm{p}} \equiv \omega_{\mathrm{p}}$ (we set $k_B=1$ from now on) is the effective temperature of ion plasma frequency. As the  typical simulated  temperatures are $T \gtrsim 10^8$ K, in our case only classical effects will be relevant.}

\subsection{Ewald sums in screened Gaussian ion systems}
To efficiently sum force and energetic contributions in our system we implement the Ewald technique \cite{Ewald} usually applied when dealing with the Coulomb potential or finite range potentials in general \cite{watanabe, aguado}. This allows an accurate evaluation of electrostatic potentials (forces) along with periodic boundary conditions (PBC). 

Briefly, Ewald decomposition is used to split the problem into real-space and Fourier-space parts as it greatly accelerates computation yielding an improved energy evaluation  \cite{holdenalphaewald}. Other alternative approaches use sums over neighboring simulation replicas \cite{Horo_2007, CaplanMulticomponent}. 

We will focus our study of OCP and MCP systems likely appearing in the outer NS crust, thus in order to illustrate typical conditions we will focus on ion densities in the range $n_I \sim 10^{-6}-10^{-4}$ $\rm fm^{-3}$, corresponding to baryon densities $n_B \sim 0.0001-0.01$ $\rm fm^{-3}$. 

{ We now describe the species simulated in our work. First, inspired by \cite{PearsonPhysRevC.83.065810} and latter updated by \cite{pearson2018} we consider a OCP with a single species $Z=38$,  $A=124$ and two MCP mixtures, see \cite{Fantina_2020}, that we label as $M_1,M_2$.} For the latter we particularly use the composition from \cite{Mckinven_2016, CaplanMulticomponent} and set five species with a common global lepton fraction $Y_e|_{M_1}\sim Y_e|_{M_2}=0.43$ and impurity parameter $Q_{imp}|_{M_1}=\sum_i X_i(Z_i-\langle Z \rangle)=21.48$  and $Q_{imp}|_{M_2}=10.69$. 

More specifically for mixture $M_1$ we take $\{Z,A,X_i\}|_{M_1}=$\{(30,69,0.407), (28,64,0.352), (42,100,0.111), (32,76,0.074), (40,96,0.056)\}. We will also consider an alternative mixture, $M_2$, differing from $M_1$ in just two ion species, in order to study the effect of the variation of the most frequent ion species in the crystallization, i.e. $\{Z,A,X_i\}|_{M_2}=$\{(32,69,0.407), (28,64,0.352), (36,100,0.111), (32,76,0.074), (40,96,0.056)\}. At this point we are aware that especially for OCP the chosen species may retain some underlying model dependence, later we will comment how our results are robust in this respect.

The crystallization is characterized by two dimensionless parameters. The first one is the Coulomb parameter $\Gamma_C=Z^{2} / l T$. The second one is the screening parameter {  $\kappa=\frac{l}{\lambda_e}\equiv k_{TF}l$. We note here that although generically $\kappa=0$ corresponds to the unscreened Coulomb case, in the relativistic theory it cannot be smaller than $\left(k_{\mathrm{TF}} l\right)_{\min }=0.185 Z^{1 / 3}$.}

\begin{figure*}[ht]
\centering{\includegraphics[width=\textwidth]{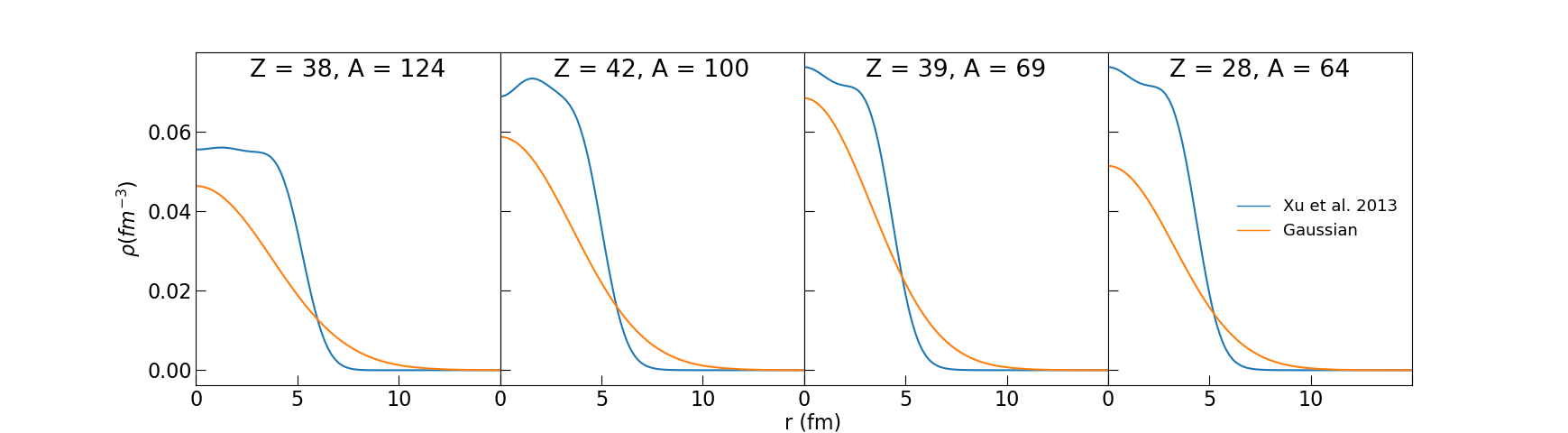}}
\caption{Charge density distribution from \cite{xuetal2013} together with Gaussian functions used in this work for four representative ions used in OCP (first left) and mixtures $M_1,M_2$. }
\label{fig2}
\end{figure*}

In order to study this system we  will be using computational  techniques, MD, to solve the equations of motion of $N_I=n_IV$ ions in a cubic box with volume $V=L^3$. Each ion is modeled as a finite-size Gaussian charge density distribution \cite{visscherdyall} in the form  $\rho_{i,a_i}(r)=Z_i\left(\frac{a_i}{\pi}\right)^{\frac{3}{2}}e^{-a_i r^2}$ where $a_i=\frac{3}{2\left\langle R^{2}\right\rangle}$ and $\sqrt{\left\langle R^{2}\right\rangle}=\left(0.8 A^{1 / 3}+2.3\right)$ reasonably describing the $A>60$ nuclear size we consider and their binding energies when compared to more refined treatments of Xu et al. \cite{xuetal2013}, see Fig.(\ref{fig2}).

In what follows we will define another dimensionless parameter, $\eta_i=1/\sqrt{a_i}l$, to characterize the charge spread of a given ith species in the OCP/MCP. This  picture thus aims to size the effect of ions when compared to previous attempts using point-like charges ($\eta=0$) \cite{watanabe, horo_2005}. We must note that previous works, see \cite{potekhinchabrier_PhysRevE.62.8554}, partially incorporated ion-ion correlations in the periodic point-like ion arrays using the structure factor of the particular lattice $S(q)$ and fitting the $q\rightarrow 0$ behaviour. They used, in the elastic part of $S(q)$ the point-proton form factor given by the Debye-Waller approximation where in the classical limit $T\gg T_p$,  $e^{-W(q, \Gamma,0)}\approx e^{-\langle r^2  \rangle q^2/6}$. This approximation mitigates the unphysical nuclear point-like nature although does not fully incorporate the refinement due to the proton charge form factor nor the Tassie-Barker correction, see \cite{DebyeWaller} for a discussion. As mentioned, this treatment somewhat corrects the point-like behaviour being its  usability limited to small momentum, $q$, for nuclear radii, $R$, fulfilling $qR \ll 1$. 

In our calculation, our simulations incorporate the ion Gaussian charge distribution description for each species  as the source of the screened fields (see Eq.(\ref{phi2})) creating the potentials and forces in the dynamical equations being solved in real space-time.

Arising from this treatment, the new resulting ionic potential is thus a superposition of those from individual screened Gaussian ions $\rho_{i,a_i} (r)$ so that when solving the Poisson equation we must replace Eq.(\ref{phi1}) by 
\small
\begin{multline}
    \phi_{Zi,a_i}\left(\vec{r}\right)=\frac{Z_i}{2|\vec{r}-\vec{r_i}|} e^{\frac{1}{4 a_i \lambda_e^2}}\left[e^{-\frac{|\vec{r}-
\vec{r_i}|}{\lambda_e}}\mathrm{erfc}\left(\frac{1}{2\sqrt{a_i}\lambda_e} - \right.\right.\\ 
\left.\left. \sqrt{a_i}|\vec{r}-\vec{r_i}|\right)  -
e^{\frac{|\vec{r}-\vec{r_i}|}{\lambda_e}}\mathrm{erfc}\left(\frac{1}{2\sqrt{a_i}\lambda_e}+\sqrt{a_i}|\vec{r}-
\vec{r_i}|\right) \right],
\label{phi2}
\end{multline}
\normalsize
with erfc the complementary error function.

In order to implement the screened interaction from Eq.(\ref{phi2}) using the Ewald summation technique \cite{Ewald} we must introduce spurious {\it screening charges} with opposite sign, $-Z_i$ to screen the real ones, $+Z_i$, in the form  $\rho_{-Z_i,\alpha_{\mathrm{Ewald}}} = -Z_i\left(\frac{\alpha_{\mathrm{Ewald}}}{\pi}\right)^{
\frac{3}{2}}e^{-\alpha_{\mathrm{Ewald}} r^2}$ with $\alpha_{\rm Ewald}$ a characteristic width parameter. To maintain the electrical charge neutrality  compensating charges $\rho_{Z_i,\alpha_{\mathrm{Ewald}}}$  must be also considered. 

The interaction energy is thus efficiently obtained from fast converging contributions of short and long-range terms, minus an extra term to exclude self-interactions and properly setting a meaningful ground state value as explained in \cite{watanabe}.
%
%
Contributions can be written as $U=U_{\mathrm{short-range}}+  U_{\mathrm{long-range}}-  U_{\mathrm{self}}$ and we detail them in what follows. 

This novel calculation for potentials and forces in the case of Gaussian ions in screened OCP and MCP is described below. First, a  short-range part describes  the interaction between the real charges and the potential created by the sum of real plus screening charges, $\phi_{\rm{short-range},i}\left(\vec{r}\right)=    \phi_{Zi,a_i}\left(\vec{r}\right)+    \phi_{-Z_i,\alpha_{\rm Ewald}}\left(\vec{r}\right)$. 

It is given by the integral expression
\begin{multline}
    U_{\mathrm{short-range}}=\frac{1}{2}\sum_{i=1}^{N_I}\sum_{j\neq i=1}^{N_I} 2 Z_j\left(\frac{a_j}{\pi}\right)^{\frac{1}{2}} \frac{e^{-a_j r_{ij}^2}}{r_{ij}}\times \\ \left[\int_{0}^{\infty} r' \phi_{\rm{short-range},i}\left(r'\right)e^{-a r'^{2}}\mathrm{sinh} \left(2a_jr_{ij}r'\right)dr'\right] \\
 -\frac{2\pi}{V} \sum_{i=1}^{N_I}\sum_{j=1}^{N_I} Z_j \int_0^{\infty} r'^{2}\phi_{\rm{short-range}, \mathrm{i}} \left(r'\right)dr',
 \label{Ushort}
\end{multline}
where $r_{ij}=|\vec{r_i}-\vec{r_j}|$ is the distance between the $ij$ particles. This contribution has a dependence in the interparticle distance that tails off quickly, so that it converges very rapidly in real space. 



A long-range part of the interaction is created by the compensating charges and the background average charge,  $\rho_{\text {avg }}=\frac{\sum_{i} Z_{i}}{V}$ under the prescription  $\sum_i   \rho_{Z_i,\alpha_{\rm Ewald}}-\rho_{\text {avg }}$. Explicitly, this is done by transforming Poisson's equation from the coordinate space to the Fourier $k$-space and including a summation over reciprocal lattice vectors so that the sum converges to a finite value under the form

\begin{equation}
\phi_{\rm long-range}(\vec{r})=\sum_{j}^{N} \sum_{\vec{k} \neq 0} \frac{4 \pi Z_{j}}{V\left(k^{2}+\frac{1}{\lambda^{2}_e}\right)} e^{\frac{-k^{2}}{\alpha_{\mathrm{Ewald}}}} e^{i \vec{k}\left(\vec{r}-\overrightarrow{r_{j}}\right)}
\end{equation}

where $\vec{k}=\frac{2 \pi}{L}\left(n_{x}, n_{y}, n_{z}\right)$ and $n_{x}, n_{y}, n_{z} \in \mathbb{Z}$. 

The associated energy term, $U_{\mathrm{long-range}}$, is thus

\begin{equation}
    \begin{aligned}
    U_{\mathrm{long-range}}&= \frac{1}{2}\sum_{i,j=1}^{N_I} \sum_{\vec{k}\neq 0} \frac{4\pi Z_i Z_j}{V \left(k^2+
    \frac{1}{\lambda_e^2}\right)}\times\\ &\left[e^{\frac{-k^2}{4}\left(\frac{1}{a_i}+\frac{1}{\alpha_{\mathrm{Ewald}}}\right)} 
    e^{i \vec{k}\left(\vec{r_i}-\vec{r_j}\right)}\right].
    \label{Ulong}
    \end{aligned}
 \end{equation}
Finally, it remains to substract the interaction between the real charge and its own compensating charge as it is included spuriously in the long-range part. It is given by

\begin{equation}
    U_{\mathrm{self}}=2\pi\sum_{i=1}^{N_I} \left(\frac{a_i}{\pi}\right)^{\frac{3}{2}}Z_i \int_0^{\infty} r'^{2} \phi_{Z_i,\alpha_{\mathrm{Ewald}}}\left(r'\right) e^{-a_i r'^{2}} dr',
\end{equation}

In our approach using MD simulations the pairwise force computation  $\vec{F}_{\mathrm{ij}}$ arises from the gradient of Eqs. (\ref{Ushort}),(\ref{Ulong}) and takes the form
\begin{equation}            
    \begin{aligned}
    &\vec{F}_{\mathrm{ij},\text{short-range}} = 2\left(\frac{a_j}{\pi}\right)^{\frac{1}{2}}\frac{Z_j e^{-a_j r_{ij}^2}}{r_{ij}^2}\left(\frac{\vec{r}_{ij}}{r_{ij}}\right)\times\\
    &\left\{\int_0^{\infty} r^{\prime}\phi_{\rm{short-range},\mathrm{i}}\left(r^{\prime}\right)e^{-a_j r^{\prime 2}}\right.\\
    &\bigg.\left[\left(1+2a_j r_{ij}^2\right)\sinh\left(2a_j r_{ij}^2\right)-2a_j r_{ij}r^{\prime} \cosh{\left(2a_j r_{ij}^2\right)} \right]dr^{\prime}\bigg\}
    \end{aligned}
\end{equation}
\begin{equation}
 \begin{aligned}
    &\vec{F}_{\mathrm{i,long-range}}=\frac{1}{2}\sum_j\sum_{\vec{k}\neq 0}\frac{8 \pi Z_i Z_j}{V \left(k^2 + \frac{1}{\lambda^2}\right)} e^{\frac{-k^2}{4\alpha_{\mathrm{Ewald}}}}\times\\
    &\left[{e^{-\frac{k^2}{4a_i}}e^{i\vec{k}\cdot\left(\vec{r}_i-\vec{r}_j\right)}-e^{-\frac{k^2}{4a_j}}e^{-i\vec{k}\cdot\left(\vec{r}_i-\vec{r}_j\right)}}\right] \frac{\vec{k}}{2i}
 \end{aligned}
\end{equation}
 
\section{MD simulations of screened OCP and MCP with Gaussian charge spread}
\label{results}

We performed our simulations using a multi-core computer infrastructure along with our original code USALMD$_{GI}$ using  \texttt{Fortran+OpenMP}. We numerically solved the ionic motion while interacting via the Debye potential in Eq.(\ref{phi2})  incorporating Ewald sums and PBC in the N$_I$VT ensemble. Up to $N_I=1024$ ions were used in this work and we verified that our energy (T)  control was good to $\delta U/U\sim 10^{-5}$. When computing averages over directions in $k$-space a maximum number of $n_{k,max}>7$  was imposed as dictated by efficiency  \cite{holdenalphaewald} in the Ewald procedure, resulting in $2 n_{k,max}+1$ replicas in each Cartesian direction. 

In our simulations we either considered a screened OCP with a single ion density, $n_I$ or a mixture, $M_1,M_2$ with impurity parameter $Q_{imp}=0, 21.48 , 10.69$, respectively. The size of the simulation box was $L=V^{1/3}$. The kinetic energy (or temperature $T$) was rescaled during the time evolution until equilibrium was achieved (typically $\sim 10^6$ fm/c with variable timestep $dt\sim 10-60$ fm/c) starting from a random distribution that we anneal when needed to the requested thermodynamical conditions. It is worth mentioning that alternative procedures to maintain constant temperature, such as thermostats, are available but since they involve additional parameters associated to the heat bath we do not expect much gain from their use, see previous works \cite{watanabePhysRevC.69.055805, APGarcia2006,universe8070380} using that of Nos\'e-Hoover in nuclear systems. 
\begin{figure}[t]
\centering{\includegraphics[width=0.97\linewidth]{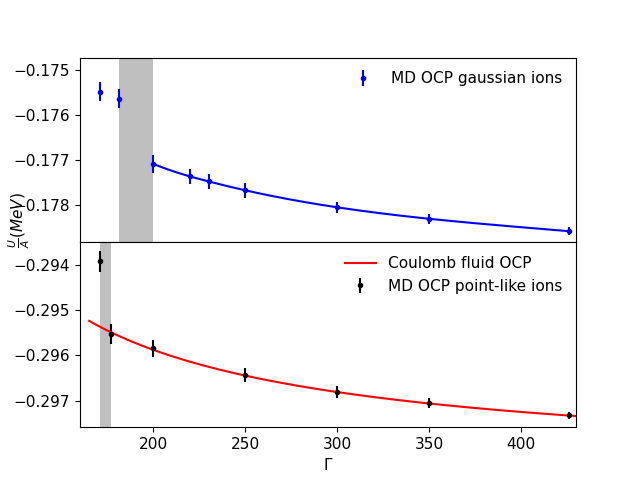}}
\caption{Potential energy per baryon as a function of the Coulomb  parameter $\Gamma$ for ions with $Z=38,A=124$ at   $n_I=2.06\times 10^{-6}\, \rm fm^{-3}$. Bottom panel: non-screened OCP as obtained  in point-like Coulomb fluids \cite{theorypollock} (solid red line) and our results from MD simulations using point-like ion distributions (black points). Upper panel: screened OCP with finite-size ion charges as obtained in MD simulations in our work (solid blue lines).{ Vertical grey bands  depict the region where our simulations predict solid-fluid phase transition and system melts.}}
\label{fig1}
\end{figure}

In Fig.(\ref{fig1}) (bottom panel) we show the energy per baryon, $U/A$, as a function of the Coulomb parameter  $\Gamma\equiv \Gamma_C$ for a screened OCP with $Z=38, A=124$ and $n_I=2.06\times 10^{-6}\,\rm  fm^{-3}$. The realistic charge distribution along with our Gaussian approximation is shown in left panel in Fig. (\ref{fig2}). { We depict with vertical grey bands the region where our simulations predict solid-fluid phase
transition and system melts. The transition involves a jump in (potential) energy for our fixed NVT ensamble. 

Let us remind here that, once the NVT  thermodynamical conditions in our MD simulation are initially fixed the equilibrated systems results in a determined solid or fluid state. Melting in the point-like Coulomb case, $\kappa=0$, is recovered at $\Gamma_m(\kappa=0)\equiv \Gamma_m \sim 175$ and accommodated in the interval we find $\Gamma_m \in [171,177]$ while that from the screened OCP we find $\Gamma_m \in [181,200]$. Previous works using Yukawa fluids \citep{khrapak}, have provided a phenomenological fit given by  $\Gamma_{m,\kappa}\equiv \Gamma_m\left(\kappa\right) = \frac{172 e^{\xi\kappa}}{1+\xi\kappa+\frac{1}{2}\xi^2 \kappa^2}$ that consistently predicts $\Gamma_{m,\kappa}=187$ for our simulated systems. From this expression, the unscreened value is predicted at $\Gamma_{m,\kappa=0}=172$, instead of 175 but this has no impact on our results as we let the ion dynamics in our simulated system evolve towards the stable final configuration.  Note that the fact that the melting parameter for our simulated screened ion system in the relativistic polarizable electron background is larger than the canonical Couloumb value does intimately depend on the ion species composition $(Z,A)$ 
and ion-electron correlation, previously shown to have no monotonous behaviour with pressure (density), see Fig. 4 in \cite{Fantina_2020}. In Fig. 3 in that same work \cite{Fantina_2020} at $T=T_m$ and $P\lesssim 5\times 10^{-7}$ $\rm MeV\,fm^{-3}$, $\langle Z \rangle \lesssim 27$, a decreasing $\Gamma_m$ tendency was shown while it was reversed afterwards up to $\langle Z \rangle \sim 42$, to jump over the canonical Coulomb value $\Gamma_m=175$ and slightly decrease from  that on.  Correspondingly, in the case we depict in Fig.(\ref{fig1}), with $Z=38$, $A=124$, we are above the $P \sim 10^{-4}$ $\rm MeV\,fm^{-3}$, $\langle Z \rangle \sim 26$ and thus the  $\Gamma_m$ that we find is larger than the canonical unscreened value.}

Our simulations nicely reproduce previous Coulomb fluid  calculations \cite{theorypollock,hamaguchi} for the point-like unscreened case using the expressions provided in Sect. \ref{sect2} with $\kappa=0$, as shown in  Fig.(\ref{fig1}) bottom panel. In the top panel we consider the screened case, obtaining a reduction up to $\sim40\%$ in potential energies resulting in a distorted melting diagram with a clear shift towards lower temperatures, i.e. higher $\Gamma$, melting parameter being predicted \cite{khrapak} around $\Gamma_m(\kappa)=187$. 
The effect of Gaussian charge distribution for this low ion density case does not cause any qualitative change, as $\eta \sim 0.11$.

\begin{figure}[t]
    \centering{\includegraphics[width=0.97\linewidth]{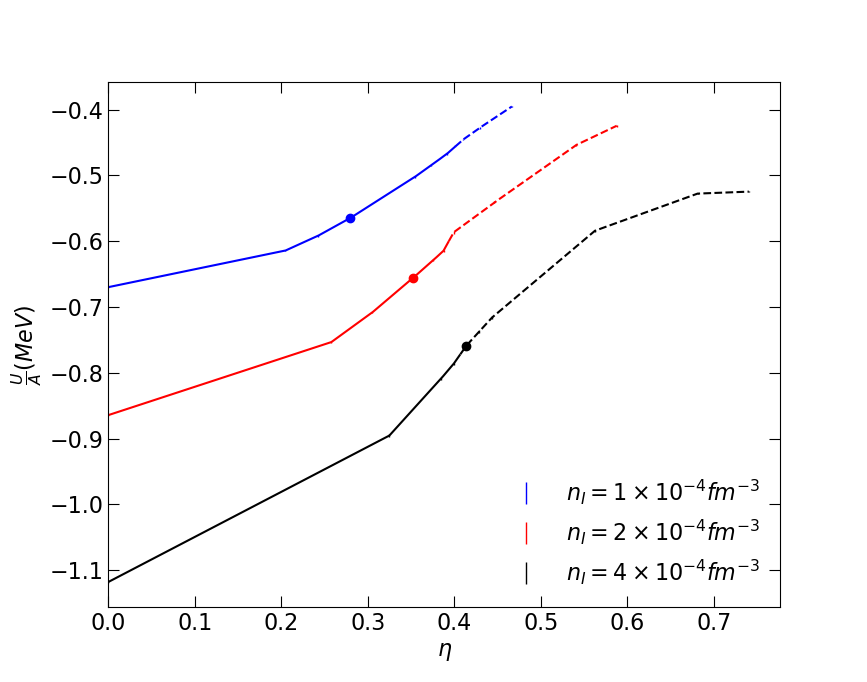}}
    \caption{Potential energy per baryon for OCP cases at $n_I=(1,2,4)\times 10^{-4} \rm fm^{-3}$, plotted against charge spread $\eta$. $\eta=0$ corresponds to point-like charges. Solid lines depict the obtained lattice region while  dashed ones correspond to melted configurations as obtained in our MD simulations. Large dots indicate the $\eta$ value corresponding to the ion, $Z=38,A =124$, see left panel in Fig.(\ref{fig2}).}
    \label{fig3}
\end{figure}

In Fig.(\ref{fig3}) we show the energy per baryon, $U/A$, as a function of the charge spread parameter $\eta=1/\sqrt{a}l$ for a screened OCP ($Z=38,A =124$) and $n_I=(1,2,4)\times 10^{-4} \rm fm^{-3}$. Point-like ions correspond to $\eta=0$. We set $\Gamma=190$ and for this case melting is approximately predicted \cite{khrapak} at $\Gamma_{m,\kappa}=187$. 

In order to size the corrections introduced when varying charge spread we use the following procedure. We  fix the ion species to  $Z=38,A =124$, since by doing this we can explore the change in ion finite size (or equivalently, A) without variation of $\Gamma$. We see that as charge distributions are realistically described having a finite spread the lattice energies are monotonically less stable. 

In addition, to evaluate the robustness of our finding we vary $\eta$ indicating on each line with a large dot that corresponding to the size assigned according to our prescription for Gaussian spread (as is A and density dependent) and determine the change with respect to point-like description finding $\Delta U/U\sim 16, 24, 36\%$ for $n_I=(1,2,4)\times 10^{-4} \rm fm^{-3}$, respectively. We note that uncertainty on this prescription does not change the behaviour. 

Solid (dashed) line shows where the lattice, although less stable, manages to exist (melts) when increasing the spread for fixed $\Gamma=190>187$ i.e. above the quoted melting parameter value. Thus for increasing values of the density (yet in the outer crust) an ionic point-like description is not appropriate when describing the energetics nor the melting behaviour as it may overpredict the existence of periodic lattices. However for densities corresponding to values of $\eta \lesssim 0.5$ this effect is no longer critical on the melting parameter but it does affect the energetic stability of the lattice. 

In order to analyze the effect of having species contamination, possibly due to an accretion episode and further processing, we consider ion mixtures $M_1, M_2$. Note that for multicomponent systems the fraction averaged $\Gamma_{MCP} = \sum_i X_i \Gamma_i$ \citep{CaplanMulticomponent} where $\Gamma_i$ are those of Coulomb theory for each ion component with fraction $X_i$ and $\langle Z \rangle = \sum_i X_i Z_i$ is the average charge density of the system.

Thus to understand the effect of both the balance of the individual ion population weights and charge spread we have performed simulations fixing $n_I$. For mixture $M_1$ at $n_{I}=1\times10^{-4}\,\rm fm^{-3}$ we find there is a change in energy $\Delta U/U\sim11\%$ for $\Gamma_{MCP} \in [190,214]$ when considering point-like compared to finite size ions. However despite a modest correction, there is an important difference, while in the scanned range  $\Gamma_{MCP}>\Gamma_{m,\kappa}$ we would expect a crystallized sample, we find this only happens for $\Gamma_{MCP}>214$ where there is a qualitative change for the most frequent ion ($Z=30, A=69$) in $M_1$ whose $\Gamma_i>\Gamma_{m,\kappa}$. Therefore we find that crystallization is affected by the individual $\Gamma_i$ of population fractions (in the point-like or finite charge spread) it being a robust effect when increasing plasma densities.

If we now compare $M_1$ and a slightly different mixture $M_2$ (where only two ion species are different) we find that for this same density  $n_{I}=1\times10^{-4}\,\rm fm^{-3}$ at same $X_i$, $\Gamma_{MCP}=195$ and  $\Gamma_{m,\kappa}=187$ a dramatic change arises leading to lattice formation in $M_2$, with a decrease of $\sim 14\%$ in energy per ion while $M_1$ stays as a fluid, irrespective of whether we use a  point-like or Gaussian charge distribution. 

\section{Conclusions}
\label{conclude}

We have simulated finite-size ionic systems immerse in a relativistic degenerate electron background using Molecular Dynamics in a fixed NVT ensemble. With this technique we effectively solve the dynamical equations for ions having a Gaussian charge spread that depends on their mass number, $A$. We have obtained expressions for pair-wise forces appropriate to use with efficient Ewald sums. This procedure is valid for single or multiple species plasmas, i.e. OCP or MCP. In addition, potential energy $U$ is also obtained in this setting. 

We find that  for densities of interest in the outer crust of NSs, where electrons are in the degenerate relativistic regime, the static Thomas-Fermi approximation yields results in agreement with relativistic Jancovici expressions at low momentum, $k\ll 2 k_{F,e}$ and this is consistent for low density crystals. Ion species population and ion-electron correlations obtained analytically in previous works in the literature from free energy minimization find melting parameters whose trends, for the cases we analyze, are in agreement with those found. A careful  evaluation of energetic stability of crystallized systems should include not only  screened interaction by means of the screening parameter, $\kappa$, but also that sizing the charge spread, $\eta$. We improve previous works using approximations where nuclei are treated as point-like objects by considering instead their finite size being able to follow the dynamics in real time.  
We find that incorporating these refinements leads to a steady decrease of energetic stability of lattice and it may lead to melting at lower temperatures than when calculated with point-like approaches. In addition, for MCP systems special caution must be taken when using effective values of the Coulomb parameter as the fractions (or equivalently the individual $\Gamma_{i}$) play a critical role to crystallize the system. We expect that this could somewhat influence the binding energy leading to less stable configurations with possible impact on the mechanical properties derived from the stress tensor.

\section*{Acknowledgments}

We acknowledge  useful comments from A. Aguado, N. Chamel, A. Fantina.  This research has been supported by University of Salamanca, Junta de Castilla y Le\'on  SA096P20, Spanish Ministry of Science  PID2019-107778GB-100 projects. DB acknowledges support from the program Ayudas para Financiar la Contratación
Predoctoral de Personal Investigador (Orden EDU/1508/2020) funded by Consejería de Educación de la Junta de Castilla y León and European Social Fund. This work was partially supported by the computing facilities SCAYLE, IAA-CSIC, CETA-Ciemat and TITAN-University of Salamanca.

\bibliography{apssamp}

\end{document}